\documentclass[11pt,prerint]{elsarticle}
\usepackage{txfonts} 

\usepackage{epsfig}
\usepackage{bm}
\usepackage{amssymb}
\usepackage{wrapfig}
\usepackage{ulem}
\usepackage{multirow}
\usepackage{color}

\newcommand{\el}{{\cal L}}

\newcommand{\psibar}{\mbox{$\overline{\psi}$}}

\newcommand{\tldF}{\mbox{${\tilde F}$}}

\newcommand{\vbr}{\mbox{$\bm{r}$}}
\newcommand{\vk}{\mbox{$\bm{k}$}}

\newcommand{\vB}{\mbox{$\bm{B}$}}

\newcommand{\nnb}{\mbox{$\nu {\bar \nu}$}}

\begin{document}

\begin{frontmatter}

\title{$\nu {\bar \nu}$-Pair Synchrotron Emission in Neutron-Star Matter 
based on a Relativistic Quantum Approach}

\author[nubs,nao]{Tomoyuki~Maruyama}

\author[WIU,nao]{A.~Baha~Balantekin}

\author[snuv,nao]{Myung-Ki Cheoun}

\author[Beih,nao,asTky]{Toshitaka~Kajino}

\author[ND,nao]{Grant J. Mathews}

\date{\today}

\begin{abstract}
We study the {$\nu {\bar \nu}$}-pair synchrotron emission from electrons and protons 
in a relativistic quantum approach.
This process occurs only in a presence of a strong magnetic field, and it is
 considered to be one of effective processes for neutron star cooling.
In this work we calculate the luminosity of the \nnb-pairs emitted 
from neutron-star-matter with a magnetic field of about  $10^{15}$~G.
We find that the energy loss is much larger than that 
of the modified Urca process.
The \nnb-pair emission processes in strong magnetic fields is 
expected to contribute significantly to the cooling
of the magnetars. 
\end{abstract}

\begin{keyword}
Neutron-Star,  Neutrino Emission, 
Strong Magnetic Field, Relativistic Quantum Approach
\end{keyword}

\address[nubs]{College of Bioresource Sciences,
Nihon University,
Fujisawa 252-8510, Japan}

\address[nao]{National Astronomical Observatory of Japan, 2-21-1 Osawa, Mitaka, Tokyo 181-8588, Japan}

\address[WIU]{Department of Physics, University of Wisconsin, Madison,
WI 53706, USA}

\address[asTky]{Graduate School of Science, The University of Tokyo, Hongo 7-3-1, Bunkyo-ku, Tokyo 113-0033, Japan}

\address[Beih]{Beihang University, School of Physics, International Research Center for Big-Bang Cosmology and Element Genesis, Beijing 100083, China}

\address[snuv]{Department of Physics, Soongsil University, Seoul, 156-743, Korea}
\address[ND]{Center of Astrophysics, Department of Physics,
University of Notre Dame, Notre Dame, IN 46556, USA}

\end{frontmatter}

\bigskip


Magnetic fields in neutron stars play important roles
in the interpretation of many observed phenomena.
Magnetars, which are associated with super strong magnetic fields,
\cite{pac92,mag3} have properties different from  
normal neutron stars. 
Thus, phenomena related magnetars can provide
a lot of information about the physics of the magnetic field.

Magnetars emit energetic photons and are observed as 
soft gamma repeaters (SGR) and anomalous X-ray pulsars (AXPs) 
\cite{Mereghetti08}. 
Furthermore, the surface temperature of magnetars is 
$T\approx 0.28  - 0.72~$keV, 
which is larger than that of normal neutron stars 
$T \approx 0.01 - 0.15~$keV at a similar age \cite{Kaminker09}. 
Thus, the associated strong magnetic fields may play a
significant role in magnetars.

Many authors have paid attention into cooling processes of neutron stars (NS) 
because it gives important information on neutron star structure 
\cite{YP2004}. 
Neutron stars are cooled by  neutrino emission, 
and a magnetic field is expected to affect the emission mechanism largely 
because a strong magnetic field can supply energy and momentum into
the process. 

Neutrino antineutrino (\nnb)-pair emission is also an
important cooling processes in the surface region of NSs. 
Pairs can be emitted by synchrotron radiation in a strong magnetic
field \cite{Land66,Kaminker09,VPSBI95,DDSL00} and 
by bremsstrahlung through two particle collisions \cite{KPPTY99,OKY14}.

The \nnb-pair synchrotron radiation is allowed 
via $p (e^{-}) \rightarrow p (e^{-}) + \nu + {\bar \nu}$ 
only in strong magnetic fields.
Landstreet \cite{Land66} studied this process for $B \sim 10^{14}$~G 
and applied it to the cooling of white dwarfs.

In that study the magnetic field is very low, and the discontinuity
due to the Landau levels was ignored when calculating the \nnb-pair
luminosity.
It was concluded that this process is insignificant

van~Dalen et al. \cite{DDSL00} calculated the \nnb-pair emission in a strong magnetic field of $B \ge 10^{16}$~G.
In such strong magnetic fields and low temperatures, $T \le 1$~MeV, 
energy intervals between two states with
different Landau numbers are much larger than the temperature.
Hence, they treated only the spin-flip transition between states 
with the same Landau number. 

In Ref.~\cite{P2Pi-1,P2Pi-2}, we introduced Landau levels in our framework and  
calculated pion production though proton synchrotron radiation strong magnetic fields.
In that work we showed that quantum calculations
gave much larger production rates than semi-classical calculations. 

In Ref.~\cite{AxPrd} we calculated the axion production in the same way, 
and found that the transition between two states with
different Landau numbers gives significant contributions even if
the temperature is low,  $T \le 1$~keV, when the strength of the
magnetic field is large, $B=10^{15}~$G.  
In this case the energy interval between the two states is much larger than the temperature.

In the present paper, then, we apply our quantum theoretical approach to
{$\nnb$}-pair synchrotron production in strong magnetic fields
and  calculate this through the transition
between different Landau levels for electrons and protons. Only this quantum approach can exactly describe the momentum
transfer from the magnetic field.

\bigskip


We assume a uniform magnetic field along the $z$-direction,
$\vB = (0,0,B)$, and take the electro-magnetic vector potential $A^{\mu}$ to be
$A = (0, 0, x B, 0)$ at the position $\vbr \equiv (x, y, z)$.

The relativistic wave function $\psi$ is obtained
from the following Dirac equation:
\begin{equation}
\left[ \gamma_\mu \cdot (i \partial^\mu - \zeta e A^\mu - U_0 \delta_0^\mu)
- M + U_s
- \frac{e \kappa}{4 M} \sigma_{\mu \nu}
(\partial^\mu A^\nu - \partial^\nu A^\mu ) \right]
\psi_\alpha (x) = 0 ,
\label{DirEq}
\end{equation}
where $\kappa$ is the anormalous magnetic moment (AMM), 
$e$ is the elementary charge
and $\zeta =\pm 1$ is the sign of the particle charge.
$U_s$ and $U_0$ are the scalar field and  time components of the vector
field, respectively.

In our model charged particles are protons and electrons.   
The mean-fields are taken to be zero for electrons, while 
for protons they are given by  relativistic mean-field
(RMF) theory~\cite{serot97}.
The single particle energy is then written as
\begin{eqnarray}
&& E(n, p_z, s) = \sqrt{ p_z^2 + (\sqrt{2 eB n + M^{*2}}
- s e  \kappa B /2M)^2} + U_0
\label{Esig}
\end{eqnarray}
with $M^* = M - U_s$,
where $n$ is the Landau number, 
$p_z$ is a $z$-component of momentum, and $s = \pm 1$ is the spin.
The vector-field $U_0$ plays the role of shifting  the single particle energy
and does not contribute to the result of the calculation.  Hence,  
we can omit the vector field in what  follows.

The weak interaction part of the Lagrangian density is written as
\begin{eqnarray}
\el_{W} &=& G_F \psibar_\nu \gamma_\mu (1 - \gamma_5) \psi_\nu 
\sum_{\alpha}
\psibar_\alpha \gamma_\mu (c_V-c_A\gamma_5) \psi_\alpha ,
\end{eqnarray}
where $\psi_\nu$ is the neutrino field, $\psi_\alpha$ is the field of the particle $\alpha$, 
where $\alpha$ indicates the proton or electron, 
while $G_F$, $c_V$ and $c_A$ are 
the coupling constants for the weak interaction \cite{rml98}. 

By using the above wave function and interaction, we obtain 
the differential decay width of the protons and electrons  
into {\nnb}-pairs.
\begin{eqnarray*}
&&  d \Gamma (n_i, n_f) = 
 \frac{G_F^2}{2^{9} \pi^5} 
\frac{ N_{\mu \nu} L^{\mu \nu}}{|\vk_i| |\vk_f| E_i E_f}
\delta\left(P_{iz} - P_{fz} -k_{iz} - k_{fz} \right) 
\delta \left( E_i - E_f - |\vk_i| - |\vk_f| \right)
d \vk_i d \vk_f d P_{fz}.
\end{eqnarray*}
with
\begin{eqnarray}
L_{\mu \nu} &=& 
2 \left( k_{f \mu} k_{i \nu} + k_{i \mu} k_{f \nu}
- g_{\mu \nu} (k_f \cdot k_i)
+ i \varepsilon_{\mu \nu \alpha \beta} 
k_f^{\alpha} k_i^{\beta} \right) ,
\\
N_{\mu \nu} &=&  \frac{1}{4} {\rm Tr} \int d x_1 dx_2 
\tldF_f(x_1-Q_{T}/2) \rho_M (n_f,s_f,P_{fz}) \tldF_f(x_2+Q_{T}/2) 
\gamma_\mu (c_V - c_A \gamma_5) 
\nonumber \\ && \quad\quad \times
\tldF_i (x_2-Q_{T}/2) \rho_M (n_i,s_i,P_{iz}) \tldF_i (x_1+Q_{T}/2) 
\gamma_\nu (c_V - c_A \gamma_5) ,
\end{eqnarray}
where
\begin{eqnarray}
\rho_M (n, s, P_z) &=&
\left[ E \gamma_0 + \sqrt{2 eB n} \gamma^2 - p_z \gamma^3
 + M^* + (e \kappa B/2M) \Sigma_z \right]
 \nonumber \\ && \quad \times
\left[ 1 + \frac{s}{\sqrt{ 2 eB n + M^{*2}} } \left(
 e \kappa B /2M + p_z \gamma_5 \gamma_0 - E \gamma_5 \gamma^3 \right)
 \right] ,
\end{eqnarray}
and 
\begin{eqnarray}
\tldF &=&  {\rm diag} \left( h_{n}, h_{n-1},  h_{n}, h_{n-1} \right)
= h_{n}  \frac{1 + \Sigma_z }{2}
+  h_{n-1}  \frac{1 - \Sigma_z }{2}  \qquad ({\rm proton}),
\\ \tldF &=&  {\rm diag} \left( h_{n-1}, h_{n},  h_{n-1}, h_{n} \right) 
= h_{n-1}  \frac{1 + \Sigma_z }{2}
+  h_{n}  \frac{1 - \Sigma_z }{2} 
\quad ({\rm electron}) .
\end{eqnarray}
Here, $\gamma^2$ and $\gamma^3$ denote the second and third Dirac gamma matrices, 
respectively, 
$\Sigma_z \equiv {\rm diag} (1, -1, 1, -1)$, and $h_n(x)$ is the  harmonic wave function 
 with the quantum number $n$.

\bigskip

In actual calculations, we use the parameter-sets of Ref.~\cite{MHKYKTCRM14}
for the equation of state (EOS) of neutron-star matter, 
which we take to be comprised of neutrons, protons and electrons.
In this work we take the temperature to be very low, $T \ll 1~$MeV,
and use the mean-fields obtained at zero temperature.

\begin{figure}[htb]    
\begin{center}
\vspace*{1em}
\includegraphics[scale=0.59,angle=270]{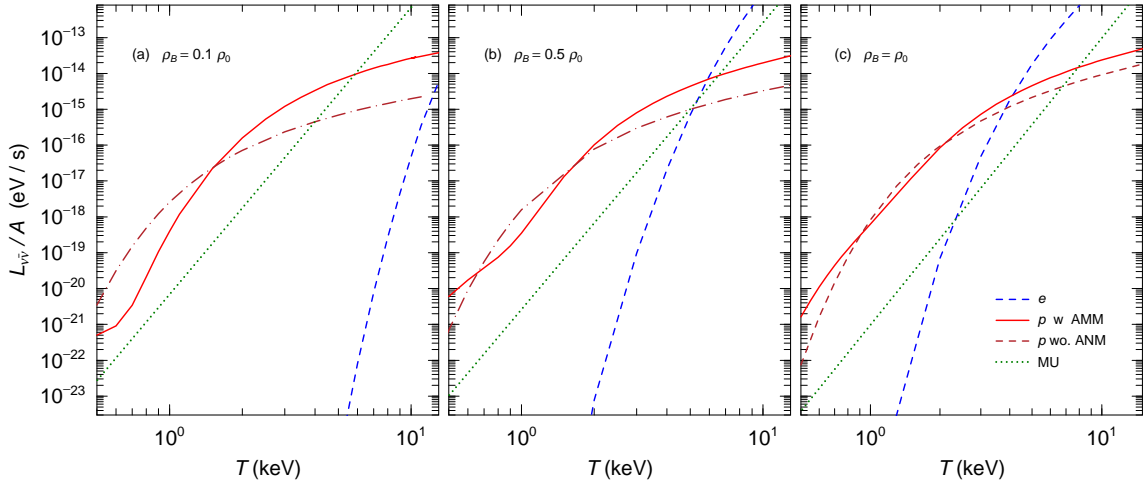}
 \caption{
$\nu {\bar \nu}$-pair emission luminosity per nucleon versus
temperature at the baryon densities $\rho_B = 0.1 \rho_0$ (a),
 $\rho_B = 0.5 \rho_0$ (a) and  $\rho_B = \rho_0$ (c)
for $B = 10^{15}$G .
The dashed line shows the contribution from electron, and
the solid and dot-dashed lines represent those from proton with
 and without the AMM, respectively.
The dotted lines indicates the results with the modified Urca process.}
 \label{figTd}
\end{center}
\end{figure}

In Fig.~\ref{figTd}, we show the temperature dependence of the neutrino
luminosity per nucleon at $B=10^{15}$G for baryon densities of 
$\rho_B = 0.1 \rho_0$ (a),  $\rho_B = 0.5 \rho_0$ (b) and  
$\rho_B = \rho_0$ (c),
where $\rho_0$ is the normal nuclear matter density.
The solid, dot-dashed and dashed lines represent the contributions from
protons with the AMM, without the AMM, and electrons, respectively.
For comparison, we show the neutrino luminosities
from  the modified Urca (MU) process \cite{MURCA} with the dotted lines.

First, we note that in the moderate temperature region, 
$T \gtrsim 3 - 5$~keV, the luminosities change rapidly 
while they vary slowly in the higher temperature region.
This qualitative behavior is very similar to
the axion luminosities in Ref.~\cite{AxPrd}.

The  energy of the {\nnb}-pair for a charged particle transition is obtained as
\begin{eqnarray}
e_{\nu} + e_{\bar \nu} 
&=& E (n_i, p_z, s_i) - E (n_f, p_z - q_z, s_f)
\nonumber \\
&=& \sqrt{2 eB n_i + p_z^2 + M^{*2}}
- \sqrt{2 eB (n_i - \Delta n_{if}) + (p_z - q_z)^2 + M^{*2}}
- \frac{e \kappa B}{M} \Delta s_{if}
\nonumber \\ &\approx&
\frac{eB}{\sqrt{2 eB n_i + p_z^2 + M^{*2}}}  \Delta n_{if}
+ \frac{p_z q_z}{\sqrt{2 eB n_i + p_z^2 + M^{*2}}}
- \frac{e \kappa B}{M} \Delta s_{if} ,
\end{eqnarray}
where $\Delta n_{if} = n_i - n_f$,  $\Delta s_{if} = (s_i - s_f)/2$,
and $n_{i,f} \gg E/\sqrt{eB}$ is assumed.

The initial and final states are near the Fermi surface in the
low temperature region,  and   $|q_z| \ll \sqrt{eB}$,
so that the energy interval of the dominant transition
is given by
\begin{equation}
e_{\nu} + e_{\bar \nu} 
\approx \Delta E = \frac{eB}{E_F^*} \Delta n_{if} 
- \frac{e \kappa B}{M} \Delta s_{if} .
\label{DelE}
\end{equation}
with $E_F^* = E_F - U_0$, where $E_F$ is the Fermi energy.

The luminosities are proportional to the Fermi distribution of
the initial state and the Pauli-blocking factor of the final state,
$f(E_i) [1-f(E_f)]$.
When $T \ll E_F^*$, the strength is concentrated in the narrow energy
region between $E_F^* - T$ and  $E_F^* + T$ for both the $E_i$ and the $E_f$.
When $T \lesssim \Delta E \approx eB/E_F^*$, however,
neither the initial nor the final states reside in the region,
$E_F^* - T \lesssim E_{i,f} \lesssim E_F^* + T$.
Then, the luminosities rapidly decrease at low temperature 
as the temperature  becomes smaller as can be seen in  Fig.~\ref{figTd}.
For example, when  $B=10^{15}$G, we find that 
$eB /E_F^* = 9.4$~keV at  $\rho_B =\rho_0$  for protons.
Indeed,  the change of the \nnb-pair luminosities
becomes  more abrupt for $T \lesssim eB/E_F^*$.

The energy step is much larger for protons than
electrons because the proton mass is much larger than the electron mass,
and the emission from protons  becomes the dominant source of \nnb-pairs.

\begin{figure}[htb]    
\vspace*{0.5em}
\begin{center}
\includegraphics[scale=0.58,angle=270]{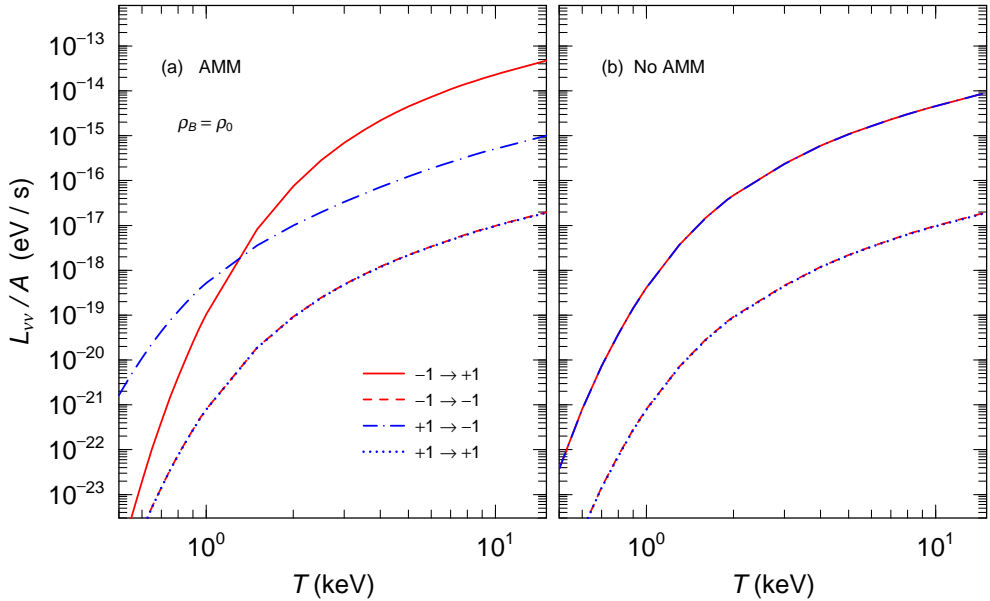}
 \caption{
Proton contribution to the {\nnb}-pair emission luminosity 
per nucleon versus temperature at a baryon density of
$\rho_B = \rho_0$ for $B = 10^{15}$G .
The left and right panels exhibit the results when the AMM is included
 and not included, respectively
The solid, dotted, dot-dashed and dotted lines represent the
 contributions when $s_i = - s_f = -1$,  $s_i = s_f = -1$, 
 $s_i = -s_f = +1$ and  $s_i = s_f = +1$, respectively.
 For the result displayed in the left panel, there are no discremable differences between the dotted and dashed lines.
 This shows that the AMM does not affect the results in the spin non-flip transition.
Similarly, in the right panel there are no differences between the dotted and dashed lines 
 and between the solid and dot-dashed lines in the right panel.}
 \label{figTEl}
\end{center}
\end{figure}

In Fig.~\ref{figTEl}, we show the initial and final spin dependence of
\nnb-pair luminosities for protons at $B=10^{15}$G and $\rho_B = \rho_0$,
when the proton AMM is included (a) and not included (b).

Without the AMM  ($\kappa_p=0$) (b), 
the contributions from the spin-flip transition,
$s_i = - s_f$,  are about 500 to 1000 times larger than those of the
spin non-flip, $s_i = s_f$.

With the AMM (a), 
 the transitions  of $s_i = -1 = - s_f$ is dominant in the higher
temperature region while the transition $s_i = +1 = - s_f$ becomes
dominant in the lower temperature region.
When $\rho_B =\rho_0$ and $B=10^{15}$G, we find that $eB /E_F^* = 9.4$~keV and 
$e \kappa B/M = 11.30$~keV. 
Thus, $eB/E_F^* < e \kappa BM$, and
 the transition with $\Delta n_{if} = n_i - n_f =2$ 
gives a dominant contribution for  $s_i = -1 = - s_f$ while 
the other transition gives the largest contribution.
For $\Delta n_{if} = 1$
the \nnb-pair production  energy in Eq.~(\ref{DelE})  is given by
$\Delta E \approx 21$~keV when $s_i = -1 = - s_f$ 
and $\Delta E \approx 7.5$~keV when $s_i = +1 = - s_f$. 
When the temperature is high enough, this positive additional energy 
$e \kappa B/M$ causes the luminosity to increase.
This sort of behavior has also
been seen in the pion production \cite{P2Pi-1}.
When the temperature is very low, however, the positive additional energy makes
the energy interval $\Delta E$ larger than the temperature, and 
it suppresses the luminosity.
The roles of the two contributions reverse at temperatures
above the inflections in Fig.~\ref{figTd}.

\begin{figure}[htb] 
\vspace*{0.5em}   
\begin{center}
\includegraphics[scale=0.63,angle=270]{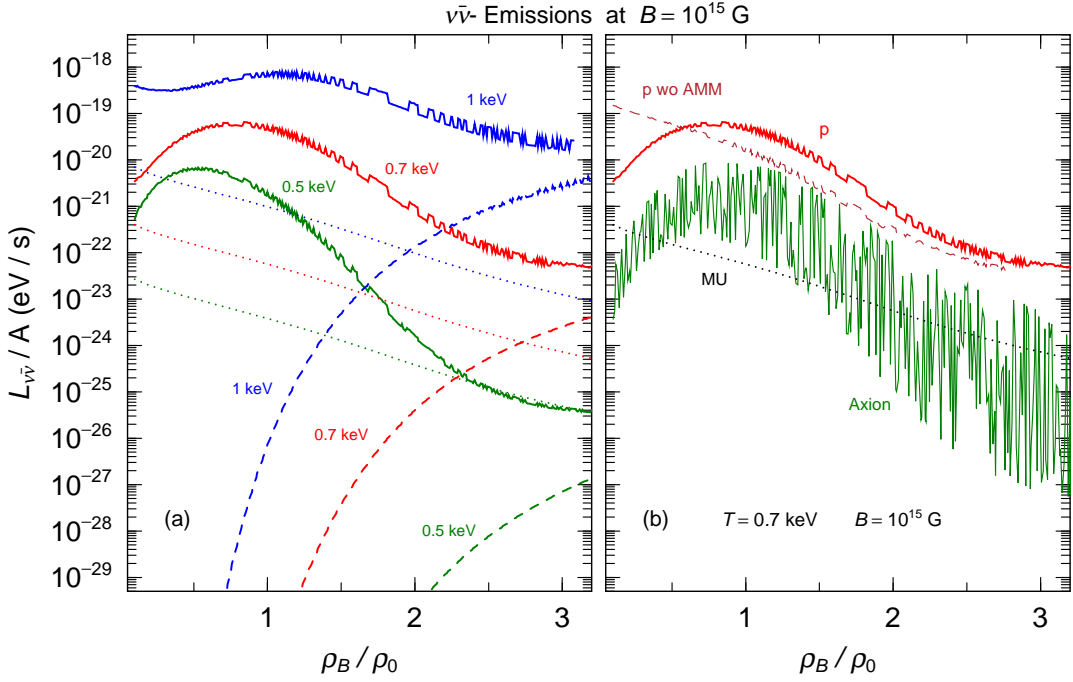}
 \caption{
Left panel: Density dependence of the {\nnb}-pair emission luminosity 
per nucleon  for $B = 10^{15}$G  at $T=0.5$, 0.7 and 1~keV 
(from bottom to top).
The solid and dashed lines represent the contributions 
from protons and electrons.
The dotted lines indicate the neutrino luminosities from the MU process.
Right panel: Luminosity per nucleon  for $B = 10^{15}$G 
at $T=0.7$~keV.
The solid and dashed lines represent the  {\nnb}-pair emission luminosity 
from protons with and without the AMM, respectively.
The thin line indicates the axion luminosity.
}
 \label{figRh}
\end{center}
\end{figure}

In Fig.~\ref{figRh}, we show the density dependence of the luminosities
at a temperature of $T=0.5$, 0.7 and 1~keV.
The solid and dashed lines represent the contributions from the protons
and electrons, respectively.
For comparison, we give the ${\bar \nu}$-luminosities 
from the modified Urca (MU) process.
In addition, we plot the proton contribution without the AMM and
the axion luminosity \cite{AxPrd} at $T=0.7$~keV on the right panel (b), 
where the strength of the luminosity is taken to be $10^{-2}$ of 
that in Ref.~\cite{AxPrd}.

The calculation results include fluctuations.
The density dependence of the factor $f(E_i)[1-f(E_f)]$ 
does not smoothly vary for strong magnetic fields and very low temperatures
because  the energy intervals between the initial and final states are 
larger than the temperature as discussed above.
However, these fluctuations are much smaller than
that of the axion luminosity because
the invariant mass of the \nnb-pair is not fixed while  
the axion mass is approximately zero.

The proton contributions are dominant at least when
$\rho_B < 3 \rho_0$.
The energy intervals for electrons are much larger than those for protons
because the electron mass is much smaller than the proton mass.
When the AMM does not exist, the electron contributions rapidly increase 
while the proton contribution gradually decreases.

At $\rho_B = 0$ and $B=10^{15}$~G, $eB/m_e \approx 11.6$~MeV for electrons, 
and $eB/M \approx 6.3$~keV for protons.
For electrons the energy interval is too large, and the transition probability
is negligibly small around zero density, while the energy interval for
protons is also large but much smaller than that for electrons, and the
proton contribution gives a finite value. 

As the density becomes larger, the Fermi energy of the electrons 
rapidly increases in the low
density region, and the electron contribution increases.
In contrast, the effective Fermi energy $E_F^*$ for protons gradually 
decreases with increasing the density 
because of the density dependence of the effective mass $M^*$, 
and  thta fact that the contribution from protons without the AMM gradually decreases.
We should note that $E_F^*$ increases in  the higher density region,
and that the proton contributions to the \nnb-luminosities must increase  
at higher densities.

When the AMM is included, we see that there are peaks in 
the proton contribution around $\rho_B \approx 0.5 - 1.2 \rho_0$.
At  $\rho_B \approx 0.7 \rho_0$, $eB / E_F^* \approx e \kappa B/M$, 
so that in the low density region, where $e \kappa B/M \ge eB/E^*_F$, 
$\Delta n_{If} \ge 2$ for the transition of $s_I = - s_f = -1$, and
$\Delta n_{If} \le -1$ for that of $s_I = - s_f = -1$ .
The density dependence of the transition ratio is different between 
the two density regions;
this change becomes more clear as the temperature decreases.
Indeed, when the AMM becomes larger, the difference is critical, and 
the neutrino luminosity shows different peaks in the two density regions.

\begin{wrapfigure}{r}{9cm}
\vspace*{-1em}
\begin{center}
  \includegraphics[scale=0.5]{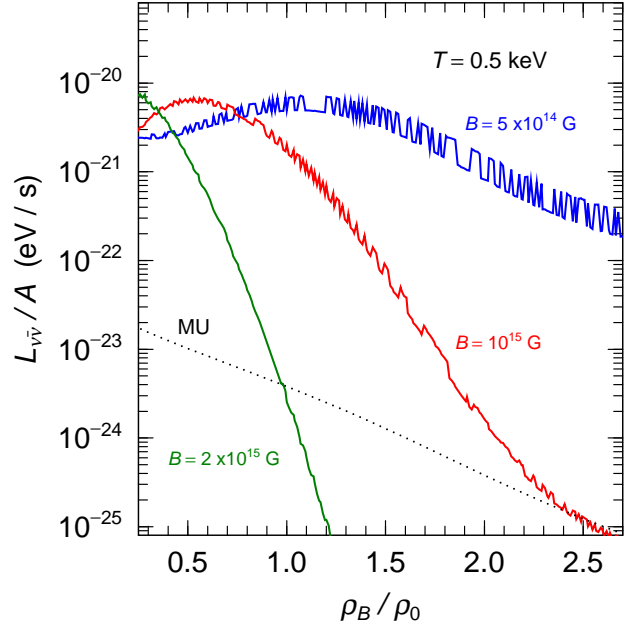}
 \caption{
Solid lines show the density dependence of the {\nnb}-pair emission luminosity 
per nucleon  for $B = 5 \times 10^{14}$~G, $10^{15}$~G and 
$2 \times 10^{15}$~G at $T=0.5$~keV.
The dotted lines indicate 
the neutrino luminosity from the MU process.
}
 \label{figRhBd}
\end{center}
\end{wrapfigure}

In Fig.~\ref{figRhBd} we show the density dependence of the
{\nnb}-luminosities at $B = 5 \times 10^{14}$~G,
$B =10^{15}$~G and $B = 2 \times 10^{15}$~G.
We see that as the magnetic field strength increases,
the luminosity decreases in the density region, $\rho_B / \rho_0 \gtrsim 1$.

As the magnetic field strength increases, the momentum transfer
from the magnetic field becomes larger, and the energy interval between
the initial and final states is also larger.
The former effect enhances the emission rate, but 
the latter effect suppresses it.
In the region of the magnetic field for magnetars, the latter effect is larger. 
Indeed, the axion production is largest around $B=10^{14}$G  \cite{AxPrd}.

Thus, the \nnb-pair emission process has a much larger effect than 
that of the MU process in strong magnetic fields.
We can conclude that  the \nnb-pair emission process is dominant
in the low density region, $\rho_B \lesssim \rho_0$, for 
a cooling process of magnetars whose magnetic field strength is 
$10^{14} - 10^{15}$~G. 
In the high density region, $\rho_B \gtrsim 3 \rho_0$, the direct Urca
process must appear, and its contribution is much larger than 
that of the \nnb-pair emission.

In summary, we have studied the \nnb-pair emission from neutron-star
matter with a strong magnetic field, $B \approx 10^{15}$~G,  
in a relativistic quantum approach.
We calculated the  \nnb-pair luminosities due to  the transitions of protons and
electrons between different  Landau levels.
In such strong magnetic fields the quantum calculation is necessary because
the energies of \nnb-pairs are much larger than the temperature.
In the semi-classical calculations energies of the \nnb-pairs  are assumed to be almost zero,
and the momentum transfer from the magnetic field cannot be taken into account exactly.
This would cause the neutrino energy spectra to shift to lower energies in the semi-classical calculation, 
 resulting in a much smaller total luminosity than that of the quantum calculations.  

In actual magnetars the magnetic field is weaker than $10^{15}$~G 
in the low density region, 
so that  in low density region the  \nnb-pair luminosity is expected to be 
much larger than that of  neutrinos  due to  the MU process.
Therefore, the present results suggest that one needs to introduce 
the \nnb-pair emission process when calculating the cooling rate of magnetars.

We expect that the cooling rate would increase due to the \nnb-pair emission process.
On the other hand, additional energy made by transitions between Landau levels could contribute to 
high energy part of the thermal  spectra of neutrinos and anti-neutrinos, 
which may heat the ambient gas surrounding magnetars  through absorption. 
Thus, the \nnb-pair production process may contribute to both the heating and cooling of magnetars whose surface temperature is larger than that of normal neutron stars.
More careful calculations of neutrino transport including these processes are highly desirable to quntity this speculation.

\medskip

This work was supported in part by the Grants-in-Aid for the Scientific
Research from the Ministry of Education, Science and Culture of
Japan~(JP19K03833, JP17K05459, JP16K05360, JP15H03665).
ABB is supported in part by the U.S.
National Science Foundation Grant No. PHY-1806368.
MKC is supported by the National Research Foundation of Korea (Grant Nos. NRF-2017R1E1A1A01074023, NRF-2013M7A1A1075764).
Work of GJM  supported in part by the U.S. Department of Energy under Nuclear Theory Grant DE-FG02-95-ER40934.

\end{document}